\def\year{2023}\relax
\documentclass[letterpaper]{article}
\usepackage{aaai20}
\usepackage{times}
\usepackage{helvet}
\usepackage{courier}
\usepackage[hyphens]{url}
\usepackage{graphicx}
\usepackage{verbatim}
\usepackage{subcaption}
\usepackage{caption}
\usepackage{graphicx}
\usepackage{booktabs}

\usepackage{graphicx}
\graphicspath{ {\string./figures/pdf/} }
\usepackage{float}
\usepackage[show]{chato-notes}
\usepackage{bbm}
\usepackage{booktabs}
\usepackage{epsfig}
\usepackage[linesnumbered,ruled,vlined]{algorithm2e} %
\SetKwInput{KwInput}{Input}                %
\SetKwInput{KwOutput}{Output}              %

\usepackage{amsfonts}
\usepackage{algpseudocode}
\usepackage{subcaption}
\usepackage{graphicx}
\usepackage{url}
\usepackage{tikz}
\usetikzlibrary{bayesnet}
\usepackage{type1cm} %
\usepackage{graphicx} %
\usepackage{xspace} %
\usepackage{balance} %
\usepackage{multirow} %
\usepackage[font={bf}, tableposition=top]{caption} %
\usepackage{bold-extra} %
\usepackage{siunitx} %
\usepackage{microtype} %
\usepackage{xfrac} %
\usepackage{mathtools} %
\usepackage[capitalise]{cleveref} %
\usepackage{hyphenat} %

\newcommand{\citet}[1]{\citeauthor{#1} \shortcite{#1}}
\newcommand{\citep}{\cite}

\newcommand{\rcmv}{\texttt{r/ChangeMyView}\xspace}
\newcommand{\ar}{\texttt{r/AskReddit}\xspace}
\newcommand{\rcmvs}{\texttt{CMV}\xspace}
\newcommand{\cmvdelta}{\ensuremath{\Delta}\xspace}
\newcommand{\cmvnodelta}{\ensuremath{\lnot\Delta}\xspace}

\newenvironment{squishlist}
{\begin{list}{$\bullet$}
 {\setlength{\itemsep}{0pt}
     \setlength{\parsep}{3pt}
     \setlength{\topsep}{3pt}
     \setlength{\partopsep}{0pt}
     \setlength{\leftmargin}{1.5em}
     \setlength{\labelwidth}{1em}
     \setlength{\labelsep}{0.5em} } }
{\end{list}}

\title{\mbox{On the Relation Between Opinion Change and Information Consumption on Reddit}}

\author{ %
  Flavio Petruzzellis,\textsuperscript{1} %
  Corrado Monti,\textsuperscript{2} %
  Gianmarco De Francisci Morales,\textsuperscript{2} %
  Francesco Bonchi\textsuperscript{2}\\ %
  \textsuperscript{1}Università degli Studi di Padova, Padova, Italy\\ %
  \textsuperscript{2}CENTAI, Turin, Italy\\ %
  flavio.petruzzellis@studenti.unipd.it, corrado.monti@centai.eu, %
  gdfm@acm.org, bonchi@centai.eu
}

\begin{document}

\maketitle \sloppy

\begin{abstract}
 
While much attention has been devoted to the causes of opinion change, little is known about its consequences. Our study moves a first step in this direction, by shedding a light on the relationship between one user's opinion change episode and subsequent behavioral change on an online social media, Reddit.
In particular, we seize an opportunity provided by the subreddit \rcmv, an online community dedicated to debating one's own opinions on a wide array of topics. Interestingly, this forum adopts a well-codified schema for explicitly reporting when an opinion change has occurred.
 Starting from this ground truth, we analyze changes in future online information consumption behavior that arise after a self-reported change in one's opinion;
and in particular, operationalized in this work as the participation to sociopolitical subreddits.
Such participation ``profile'' is important as it represents one's information diet, and is a reliable proxy for, e.g., political affiliation or health choices.

We find that people who report an opinion change are significantly more likely to change their future participation in a specific subset of online communities.
We characterize which communities are more likely to be abandoned after opinion change,
and find a significant association (Pearson $r=0.46$ ) between propaganda-like language used in a community and the increase in chances of leaving it.
We find comparable results (Person $r=0.39$) for the opposite direction, i.e., joining these same communities.
We also show that the textual content of the discussion associated with opinion change is indicative of which communities are going to be subject to a participation change.
In fact, a predictive model based only on the opinion change post is able to pinpoint these communities with an average precision@5 of 0.20, similar to what can be reached by using all the past history of participation in communities.%

Our results establish a link between opinion change and online information consumption.
Moreover, they shed light on the role of online propagandistic communities as a first gateway to internalize a shift in one's sociopolitical opinion.

\end{abstract}

\section{Introduction}
\label{sec:intro}

Opinion dynamics is the field of study that deals with how people's opinions form and evolve in a social context~\citep{french1956formal,harary1959criterion}.
This branch of social psychology has received growing attention, also due to the widespread adoption of social-media platforms and their effect on opinion formation.
Most of the effort has been devoted to understanding the \emph{causes} of opinion changes,
by analyzing the dynamics of the opinion formation process, simulated by means of 
\emph{agent-based models}~\citep{degroot1974reaching,friedkin1990social,axelrod1997dissemination,deffuant2000mixing};
although recently, studies analyzing real-world social-media data~\citep{xiong2014opinion,tan2016winning,monti2020learning,naskar2020emotion} are emerging. While empirical studies of opinion change are lacking due to the complexity of measuring such a construct~\citep{flache2017models}, studying the \emph{externalities} of opinion change is even more challenging, for several reasons.
First, the effects may be delayed in time, which complicates the data-gathering process.
Second, while significant, the effects might be small and hard to capture, which requires large amounts of data to tease them out.
Third, there may be several confounders to the analysis.
For all these reasons, limited attention has been devoted to the consequences of opinion change. 

In this work, we move a first step towards filling this gap.  We do so by seizing an opportunity provided by the subreddit \rcmv, an online community dedicated to debating one’s own opinions on a wide array of topics. A user posts an opinion they hold, and other people in the community argue against it in the comments, by providing alternative viewpoints.
Most importantly, the original poster explicitly marks which comments succeeded in changing their view, by awarding them a \emph{delta}~(\cmvdelta).
These features makes \rcmv a natural testbed for testing hypotheses about opinion dynamics.
The main question we tackle in this paper is the following: \emph{how does opinion change relate to behavioral change?}
In particular, the behavioral change we focus on is in \emph{information consumption}, measured as participation to a set of sociopolitical subreddits.
Such participation ``profile'' is important as it represents one's information diet, and is a reliable proxy for, e.g., political affiliation~\cite{defrancisci2021no} or health choices~\cite{kumar2021covid}.

To answer this question we conjoin data from \rcmv (\rcmvs) with user activity on a set of politically-charged subreddits. First, we design a model to predict future changes in a user's participation to subreddits based on that user's submission on \rcmvs.
Doing so, we aim to answer the more specific question: is there a relationship between the change of opinion and change of behavior in a user?
We find that by using only information about the topic of that single submission, it is possible to achieve a precision similar to what can be reached by using information about past history of participation in communities, which is more notably and intuitively related to the future activity of a user~\cite{massachs2020roots}. We then focus our analysis on characterizing those subreddits which are more likely to see a change in participation after an opinion change. 
In this case, we ask wether there are specific communities which are more likely to be subject to behavior change.
By measuring the increase in the odds of a participation change after an opinion change, we find a statistically significant effect for a handful of subreddits.
In other words, some communities are more likely to be abandoned or joined after a user reports a change in their opinion.
For instance, the subreddit of Donald Trump's supporters \texttt{r/The\_Donald} is roughly $20\%$ more likely to be abandoned after a user reports an opinion change.
Finally, we ask how can these communities be characterized by looking at their language.
We quantitatively show that there is a correlation between the usage of language typical of propaganda and satire and how much the participation in the subreddit is sensitive to opinion change.

Our results suggest that opinion change is indeed consequential, and has a measurable effect on behavioral patterns within the online platform.
It is known that online behaviors are predictive of other offline behaviors~\cite{yamamoto2015social}.
Therefore, understanding opinion change online is a data-rich task with important implications, such as generating valuable proxies for other social constructs (e.g., political participation).
The present study thus represents an important validation for the field of opinion dynamics.

\section{Related Work}
\label{sec:related}

In recent years, Reddit's \rcmv has consistently been used as a data source for empirical studies about online debate and persuasion dynamics.
On the one hand, studies have focused on the structure of online debates, which are inherently different from offline ones, since they are often asynchronous and anonymous. 
\citet{ocal2021reasoning} analyse a relatively small sample of manually annotated discussions on \rcmvs to study reasoning processes in online debates.
They identify the most frequent patterns used by users to provide evidence to support their arguments in online discussions. They find that personal experience is often provided as supporting evidence for arguments, even when people claim to be experts in a field, and especially when they are discussing domestic politics.
The corpus-based study by \citet{musi2018how} uses \rcmvs discussions to enrich the theoretical standpoint that identifies concessions as one of the most effective strategies to change people's mind in a discussion.
The findings of the study suggest that concessions in online debates mainly serve two purposes: to prevent counterarguments and to introduce them.

A different perspective is taken by studies about the effectiveness of peoples argumentative strategies.
\citet{tan2016winning} analyze a corpus of discussions on \rcmvs to study the dynamics of persuasive discourse online.
Their findings highlight the importance of the formal features of discussions: independently of the topic, the language used to present one's arguments and the dynamics of interaction between the users involved in the debate prove to be important features in the task of predicting the successful change of perspective of the author.
\citet{priniski2018attitude} focus on online debate specifically regarding socio-moral issues, testing the hypothesis that it is inherently harder to change one's mind about such topics. Their findings suggest that providing factual evidence supporting one's arguments is useful to change people's mind, but not more so when debating socio-moral topics, despite the fact that factual evidence is provided more often in debates about these topics.

To the best of our knowledge, this is the first study that employs data from \rcmv discussions to provide empirical evidence about the impact of online debates on users online behavior.

\section{Data}
\label{sec:data}

Reddit is a discussion forum organized in topical communities, called subreddits.
All users must have a pseudonymous account in order to participate.
Users can post submissions in these subreddits, and comment on other submissions and comments, thus creating a tree structure for the overall discussion.
In addition, users can also upvote a submission to show approval, appreciation, or agreement (and their opposites with a downvote).
The score of a submission is the number of positive votes minus the number of negative votes it has received.
Differently from other social media like Facebook or Twitter, Reddit's homepage is organized around subreddits, and not on user-to-user relationships.
As such, the subreddits chosen by a user represent the main source of the information they consume on the website.

The subreddit \rcmv is a community on Reddit dedicated to challenging one's own opinions, with over $1.3$M subscribers at the time of writing.
Specifically, posters offer to debate their stance on a particular topic, and commenters respond by trying to change the poster's opinion.
When the poster acknowledges that a comment has succeeded in changing their opinion---even partially---they leave a comment with the codified symbol `$\Delta$'.
The presence of such labelling makes this forum an ideal place to study opinion change.
We collect public data from this subreddit from the Pushift collection~\citep{baumgartner2020pushshift}, for a 7-year time period starting from 2013, which results in \num{39845} posts by \num{26074} authors with \num{2422379} comments on them.

We are interested in understanding whether this declared change in opinion is related to a change in information consumption behavior in the future.
Previous work~\cite{massachs2020roots,balsamo2019firsthand} has showed that participation in subreddits is a valuable proxy for a wide range of characteristics of the user.
For this reason, we take the subreddits they participate in---their communities on Reddit--- as a meaningful measure of the behavior of the user.
In particular, participating in a subreddit affects a user's information diet: by subscribing to a new subreddit, a user sees its posts on their Reddit homepage.
In order to reduce the spread of effect that might occur due to different topics of discussions, we limit our scope to \emph{sociopolitical} subreddits.
We follow the definition of sociopolitical given by \citet{moy2006predicting}, thus including topics related to political figures, parties, institutions, and cultural, social, or national issues.
To select this subset of subreddits, we manually label the \num{2000} most popular subreddits in 2019 as either sociopolitical or not; we do so by looking at their description, and a sample of their posts.
This way, we find 51 such sociopolitical subreddits (see Appendix \ref{sec:appendix}).

To keep the topic of opinion change events we analyze consistent with the sociopolitical theme, we restrict our analysis to posts on \rcmvs with sociopolitical content.
To do so, we develop a supervised post classifier based on a sample of posts from the subreddits we labeled as sociopolitical.
Specifically, we build our dataset by taking a random sample of $50$ posts per month for each sociopolitical subreddit, and a sample of equal size from the others we analyzed and labeled as non-sociopolitical.
The model used is a logistic regression with L1 regularization, and the features are simple n-gram counts ($n \in [1,3]$).
The classifier obtains an average F1 score of $89.5\%$ on a held-out test set.
We also build a set of $200$ posts from \rcmvs that we manually label as sociopolitical or not: on this validation set, the classifier obtains an F1 score of $82\%$.
We then proceed to use this classifier on all \rcmvs posts.
Out of the $65$k \rcmvs posts with textual content, we identify $46$k as sociopolitical.

Then, for each \rcmvs user who authored a sociopolitical post, we gather the set of subreddits they participate in.
We define participation as writing at least one comment in that subreddit in a given time period, which is a strict proxy for having joined the community as it excludes `lurkers'.
In order to answer our research questions, we separate participation \emph{before} and \emph{after} a user's post on \rcmv.
Furthermore, we limit the time frame in which we observe behavioral change by considering only comments written at most one year before or after the date of the user's post on \rcmvs.
We do so in order to consider in our analyses only events of online behavior change that are relatively close in time to the reported opinion change, and that can thus be considered related to it more reliably.
We indicate with $b_{u, s}$ and $a_{u, s}$ the number of comments authored by user $u$ on subreddit $s$ before and after (respectively) their post on \rcmv, limited to the time frame described above.
Using these values, we define:

\smallskip
\begin{squishlist}
	\item $L_u$ as the set of subreddits that user $u$ \emph{left} after their post on \rcmv; that is, $L_u = \{ s \mid b_{u, s} > 0 \wedge a_{u, s} = 0 \}$.

\smallskip
	\item $J_u$ as the set of subreddits that user $u$ \emph{joined} after their post on \rcmv; that is, $J_u = \{ s \mid b_{u, s} = 0 \wedge a_{u, s} > 0 \}$.

\smallskip
	\item $C_u$ as the set of subreddits that \emph{changed} for user u after their post on \rcmv, that is $C_u = L_u \cup J_u$.

\smallskip
	\item $S_u$ as the set of subreddits that user $u$ \emph{stayed on} after their post on \rcmv; that is, $S_u = \{ s \mid b_{u, s} > 0 \wedge a_{u, s} > 0 \}$.
\end{squishlist}

We gather this data for all the \rcmv posts and their authors.
By using this data we define an activity profile associated to a post as a vector $Z$ that indicates, for each sociopolitical subreddit, whether the author of the post \emph{joined}, \emph{left}, or \emph{stayed} in that subreddit.
Finally, we further filter our dataset by considering only one post per user, in order to consider the user (rather than the post) as our statistical unit.
If a user has assigned more than one delta on \rcmvs, we consider the closest one to the midpoint in the dataset's timespan, in order to have the best possible conditions to build the user profile after applying the one-year filter on the time of comments.
This decision does not alter the dataset significantly, as $84.1\%$ of authors have assigned only a single delta.
The final dataset contains $\approx29$k users, and statistics about it are reported in \Cref{tab:data-stats}.

\subsection{Textual corpus}
\label{subsec:textual-corpus}
We use two sources of textual information in our analyses: the text of the submissions of users on \rcmvs and a sample of the textual contents of the sociopolitical subreddits which we consider.
We apply the same pipeline of text preprocessing to both sources:
we strip URLs, include only words that are written with latin alphabet and are at least 2 characters long, then lemmatize by using the \texttt{Python} library \texttt{nltk}~\citep{bird2009natural}.
We use this preprocessed textual data to build a tf-idf representation of both posts and subreddits.
We use the \texttt{Python} library \texttt{Scikit-Learn} \citep{pedregosa2011scikit}, filter stopwords, include unigrams and bigrams, and limit the number of features to lemmas having a minimum document frequency of $10$, which results in a vocabulary of \num{85078} lemmas.

\begin{table}[tb]
	\caption{Dataset statistics for the %
	subsets of \rcmvs posts with (\cmvdelta) and without (\cmvnodelta) a delta:
	number of users $|U|$, number of posts $|P|$, average ($\mathbb{E}$) and standard deviation ($\sigma$) of number of non-zero elements in user profiles $|Z|$.}
	\label{tab:data-stats}
	\centering
	\begin{tabular}{lrrrr}
\toprule
User set &  $|U|$ &  $|P|$ &  $\mathbb{E}(|Z|)$ &  $\sigma(|Z|)$ \\
\midrule
\rcmvs \cmvdelta	&     \num{13871} &     \num{18843} &                   \num{4331} &                   \num{4434} \\
\rcmvs $\cmvnodelta$ &     \num{15026} &     \num{21005} &                   \num{4713} &                   \num{4687} \\
\bottomrule
\end{tabular}

\vspace{-3mm}
\end{table}

\begin{table}[tb]
	\caption{Comparison between activity on sociopolitical (\textsc{sp}) and other subreddits for different groups of users: participants in \rcmv (\rcmvs), participants in \ar (\texttt{AR}), users who self-report opinion change (\cmvdelta), and users who do not (\cmvnodelta). Values represent the average over considered users. }
	\label{tab:stats-user-activity}
	\centering
	\begin{tabular}{lrrrrr}
\toprule
  User set 	&  Number of & \multicolumn{2}{c}{Comments}   &  	\multicolumn{2}{c}{Score} \\
	\cmidrule(lr){3-4} \cmidrule(lr){5-6}
	& subreddits	&  \textsc{sp} 	  & 	others   &  	\textsc{sp}  		&  others \\
\midrule
	\rcmvs	           	&         \num{107.3}  &     \num{418.8}   &     \num{2460.0}  &      \num{8.9} 	&   \num{7.4} \\
	\texttt{AR}            	&         \num{45.6} &     \num{94.5} &     \num{663.4} &      \num{10.9} 	&   \num{8.1} \\
\midrule
	\rcmvs \cmvdelta		&     \num{109.8} &     \num{402.2}  &        \num{2562.4}  &      \num{9.6} 	&   \num{7.8} \\
	\rcmvs \cmvnodelta	&     \num{104.4} &     \num{438.5}  &        \num{2341.9}  &      \num{8.2} 	&   \num{7.0} \\
\bottomrule
\end{tabular}

\end{table}

\subsection{Characterizing \rcmv}
\label{sec:cmv-vs-ar}
To better understand this dataset, as a preliminary analysis we compare the behavior of \rcmv participants to participants in a generalist subreddit, \ar, the most popular board on the site.
In this subreddit, users ask and answer questions on any topic, without any intent to challenge their opinions.
These characteristics make it is a suitable choice as a control group.
We compare these two groups in terms of activity, breadth of interests, community feedback, on the selected sociopolitical subreddits and on the rest.

\Cref{tab:stats-user-activity} reports this comparison (computed on a 10\% random sample of \ar users for efficiency).
The set of \rcmvs users is more active on the platform, both in terms of number of comments, and breadth of interests, with more than double the number of participated subreddits.
This difference is to be expected as \ar is a natural landing point for new users, while \rcmvs users are more engaged with the platform.
The fraction of activity on sociopolitical subreddits is slightly higher for \rcmvs (14.5\%) than for the others (12.5\%).
This result corroborates our choice of sociopolitical subreddits, and suggests that our set of \rcmvs users is indeed more interested in sociopolitical topics.
Instead, the community feedback, as measured by the average score of users' comments, is similar, albeit slightly lower for \rcmvs.
Finally, we observe that the distribution of such comments across the different sociopolitical subreddits follows approximately the same ranking in the two groups (Pearson $r=0.97$, not shown).
Therefore, the selected group of users does not seem to be biased towards a specific topic w.r.t. the general Reddit population.

Similarly, as a preliminary analysis, we investigate the difference in general behavior between the group of \rcmvs users who assigned a \cmvdelta and the ones who did not.
The two groups do not significantly differ in terms of average activity on Reddit, both on sociopolitical and general subreddits.
Also number of subreddits and community feedback are comparable.
However, as we show next, we are able to identify a relationship between the event of opinion change, codified by the assignment of \cmvdelta, and a change in the future activity of an individual in terms of information consumption.

\section{Results}
\label{sec:results}

We use the collected data to answer the following research questions.

\begin{squishlist}
 \item [\textbf{RQ1}:] Is there a relationship between the change of opinion and change of behavior in a user?
 \item [\textbf{RQ2}:] Are there specific communities which are more likely to be subject to behavior change?
 \item [\textbf{RQ3}:] If so, how can we characterize these communities more affected by opinion change?
\end{squishlist}

To answer these questions, we train a classifier aimed at predicting future behavior of the user by using information about its opinion change as input, and show that the two are indeed related.
Then, we identify a subset of communities that are significantly affected by opinion change.
We do so by comparing the chances of behavioral changes before and after the event of opinion change.
Finally, we characterize this subset of communities, and find that they employ a particular language, akin to propaganda and satire.

As a starting point, we compare the rate of change in user profiles between \rcmv users and a propensity-score matched one of similar size coming from \texttt{AskReddit}.
The propensity score takes into account the global user activity, the specific user activity on the 51 sociopolitical subreddits, and the timespan of the user on Reddit.
Overall, we find that users in the \rcmv group are more likely to experience behavioral change in the future ($\textrm{OR}=1.511$) than users posting on \texttt{AskReddit}.
This result highlights that a major component of opinion change, and the subsequent behavioral change, is engagement and open-mindedness by the user.

\subsection{Relationship between behavioral and opinion change}
\label{subsec:rq3}

\begin{table}
	\caption{Performance metrics for the behavior change classifier. Average and SD of the metrics across 10 cross-validation folds for the two types of input vectors.} 
	\label{tab:classifier-out}
	\centering
	\begin{tabular}{lll}
\toprule
{Input vector} &         AveP@1 &         AveP@5 \\
\midrule
Participation history &  0.258 ± 0.032 &  0.229 ± 0.015 \\
\rcmvs submission content &  0.215 ± 0.016 &  0.201 ± 0.009 \\
\bottomrule
\end{tabular}

\end{table}

\begin{figure}
	\includegraphics[width=\columnwidth]{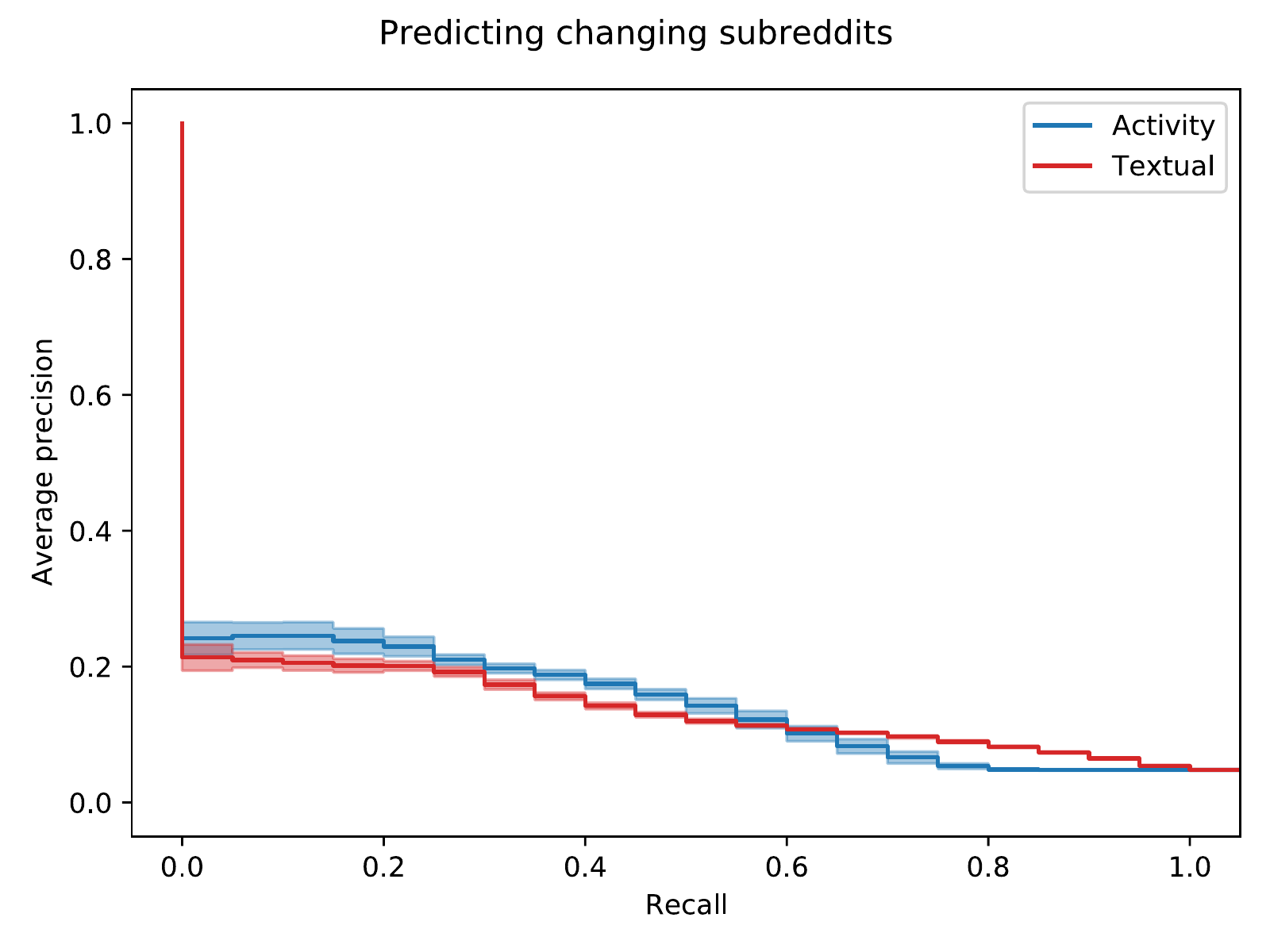}
  \caption{Precision-recall curves of two predictive models for behavior change, one using the past activity profile of a user (Activity, in blue), and another one using the textual information in the post on \rcmvs (Textual, in red).  The shaded area indicates the standard deviation of the measure across 10 cross-validation folds.}
  \label{fig:PR-curves}
\end{figure}

\begin{figure*}[t]
  \begin{center}
  	\includegraphics[width=0.6\textwidth,trim={0 0 0 18mm},clip]{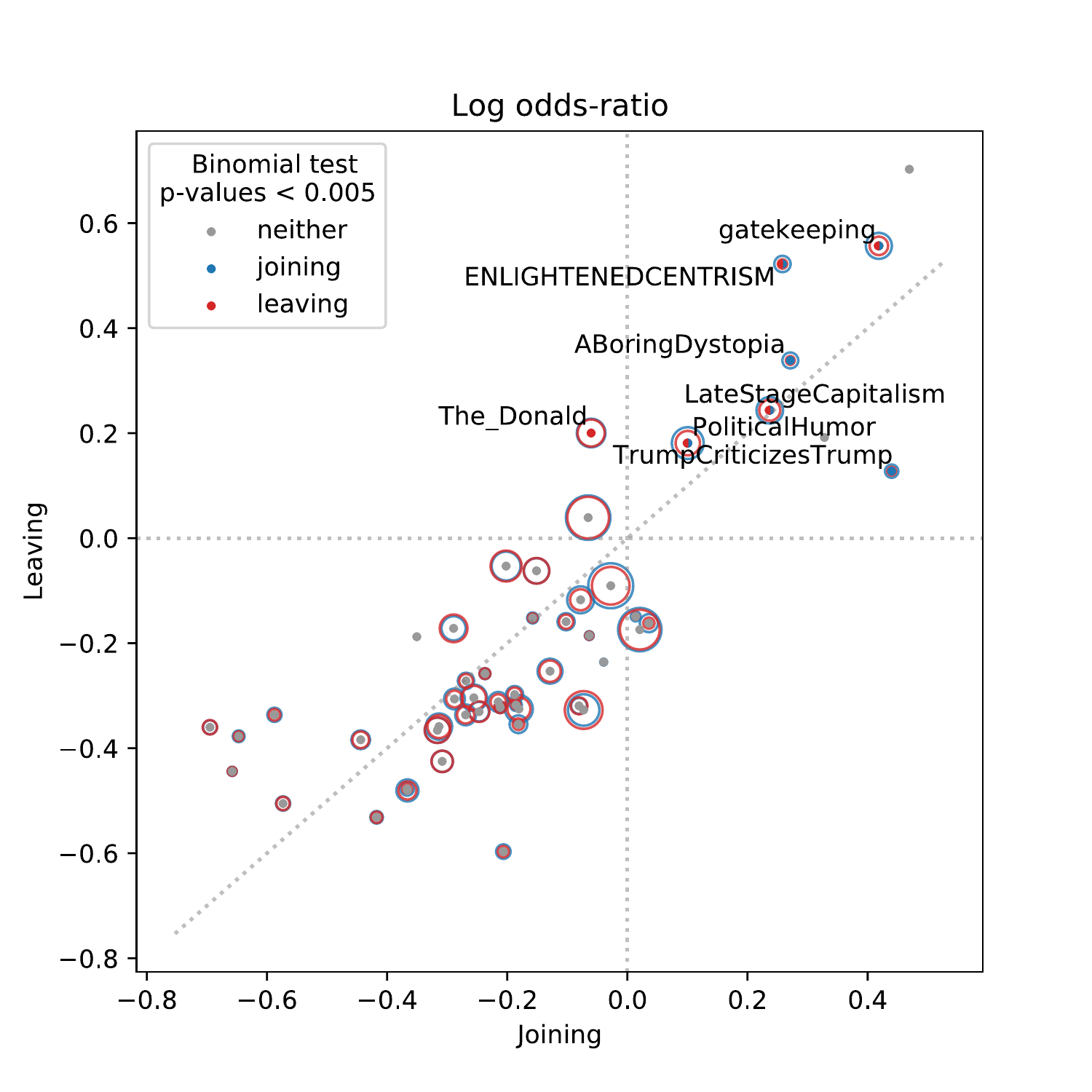}
    \caption{Log odds ratios of joining (X axis) or leaving (Y axis) each subreddit after opinion change. For each subreddit, we represent with circles area the support of their binomial test -- i.e., the number of users joining that subreddit (in blue) or leaving it (in red) after opinion change. Blue dots (respectively, red dots) represent subreddits significantly ($p < 0.005$) more likely to be joined (resp., left) after opinion change. We show the name of the subreddit for all those significant in any direction.
  	The subreddits in the top-right quadrant are those for which an opinion changed marked by a \cmvdelta is more likely to induce a behavior change.
    }
  	\label{fig:or-delta}
  \end{center}
\end{figure*}

Our first goal is to understand whether the topic of the discussion where an individual changes their minds is related to the communities that they will change in the future.
We also wish to compare the relevance of the topic of discussion against other similarly relevant pieces of information, such as the past activity of the user.
To investigate these aspects, we frame the task as a multi-class classification one.
The task consists in associating, to each author of a submission on \rcmvs in the \cmvdelta group, the subreddits for which their participation status will change in the future (i.e., pass from activity to inactivity, or vice versa).
The possible subreddits where change can happen are 51---i.e., the set of sociopolitical subreddits.
To tackle this task, we use an ad-hoc, feed-forward neural network and compare two possible input features.

The first type of input contains \emph{textual information} about the post of the user in \rcmvs. 
It is composed of two concatenated vectors: first, the tf-idf representation of a user's submission to \rcmv; second, the cosine similarity between such a vector and the tf-idf representation of each of the candidate subreddits.
These latter representations are obtained by vectorizing as a single document a random sample of submissions in a given subreddit.

The second type of input for the model contains information about \emph{past activity} of the user in sociopolitical subreddits.
Such a data source is the aforementioned user profiles $Z$: each community participated by the user is given as input to the model.
More specifically, this input contains the number of comments written by a user on each subreddit before the submission on \rcmvs.
We choose such an input since it has been shown to be predictive of future user behavior~\cite{massachs2020roots}.

For both inputs, the output layer of the neural network is composed of 51 sigmoid units. Each unit represents the probability that a specific subreddit will be left or joined by the given user after their \rcmv submission.
We test this neural network in four versions: one without any hidden layer, and with hidden layer of size 10, 100, or 1000.
We evaluate the classifier via a nested 10-fold cross-validation procedure, to perform both a hyperparameters search on the number of hidden units and an evaluation of its performance.
For each hyperparameter fold of the inner CV we select the best architecture based on its obtained Average Precision@5 (AveP@5).
The selected model is then retrained and tested on the outer CV, by computing Average Precision@1 and Average Precision@5.
In all cases, the AveP@k metrics are computed by using micro-averaging, i.e., all users in the test set are weighted the same.
 
A comparison of the results obtained by the classifier with the two inputs is reported in \Cref{fig:PR-curves} and \Cref{tab:classifier-out}.
By using only the textual representation of the single \rcmv submission, the model obtains an AveP@5 of $0.20$.
In other words, the classifier is able to list a subreddits that is going to flip after the opinion change event with 5 guesses in $20\%$ of the cases.
As a comparison, our alternative model based on the whole history of a given user---i.e., all the other communities that they participated in---obtains a similar AveP@5 of $0.23$.

We conclude therefore that there is a connection between the topic of the opinion change event, as measured by the textual representation of the submission, and some of the future behavioral changes of that user.

\subsection{Communities affected by opinion change}
\label{sec:rq-communities}

After assessing a relationship between changes in subreddit participation and the \rcmv submission,
we then investigate which changes in user behavior are more likely after opinion change.

To do so, we define $p^{J, g}_s$ as the empirical probability that a user in group $g$ will join subreddit $s$; analogously, $p^{L, g}_s$ is the empirical probability that a user in group $g$ will leave subreddit $s$.
Both empirical probabilities are built by using the users' profiles described in \Cref{sec:data}.
Hence they only take into account changes in participation occurred up to one year after the submission of the user on \rcmvs.
The two considered groups $g$ are \cmvdelta and \cmvnodelta: the first group represent individuals that have assigned a \cmvdelta, thus self-reporting that they experienced opinion change; the second group is the control group, i.e., individuals that did not assign a \cmvdelta in their \rcmvs post, and hence are considered not to have experienced opinion change on that issue.

We estimate all these probabilities from our dataset, by counting the occurrences of these events for each subreddit.
Then, we consider for each probability $p_s$ the corresponding odds \mbox{$O_s = p_s / (1-p_s)$}.
Finally, we compute the odds ratios between the \cmvdelta and \cmvnodelta groups as
\newcommand{\OR}{\mathit{OR}}
$\OR^J_s = O^{J, \Delta}_s / O^{J, \lnot\Delta}_s$
for joining a subreddit $s$.
Therefore, $\OR^J_s$ represents the increase in the odds of joining a subreddit $s$ after experiencing opinion change; $\OR^L_s$ the increase in the odds of leaving it.
We show these odds ratios for all the considered subreddits in \Cref{fig:or-delta}.
The figure uses log odds ratios to paint a symmetric view around zero (no difference between groups, i.e., $\mathrm{OR}=1$).

Interestingly, the odds ratios of joining and leaving seem to be correlated, i.e., there does not seem to be a single direction for the behavior change, and it depends on the specific user.
The only exception is \texttt{The\_Donald}, for which the effect of giving a delta seems to be clearly in the direction of leaving the subreddit (and not joining it), and which seems to coincide with  its quarantine and eventual ban from Reddit.
In addition, we find no global effect, i.e., the \cmvdelta group does not have a higher OR of leaving or joining than the \cmvnodelta group.

Then, in order to assess the statistical significance of the increase in odds, we perform a binomial test for each subreddit.
Specifically, we test whether the occurrences of a user leaving (or joining) a subreddit are significantly greater among individuals who reported opinion change with respect to those who did not.
We employ a Bonferroni correction in order to account for the large number of tests.\footnote{We remark that the Bonferroni method is the most conservative one, i.e., the chances of false positives are lower than other methods.}
We report these results with the color of the dots and the presence of the label in Figure~\ref{fig:or-delta}.
Therein we also report information about the support of each binomial test (i.e., the number of users joining that subreddit or leaving it after opinion change), represented as the size of the circle surrounding each dot.

While for most subreddits giving a delta has no or negative effects on behavior change, there are a handful of subreddits for which this effect is positive.
From the results reported in Figure~\ref{fig:or-delta}, we can in fact clearly observe two distinct groups of subreddits.
The first one is centered in the bottom-left quadrant; the log odds ratios of these subreddits are generally non-positive and there is no significant increase.
The second and more interesting group is centered in the top-right quadrant: these subreddits exhibit a significant increase in flipping one's participation to those subreddits after a self-reported change in opinion.
This group comprises $7$ subreddits: 
\texttt{gatekeeping}, 
\texttt{EnlightenedCentrism}, 
\texttt{ABoringDystopia}, 
\texttt{LateStageCapitalism}, 
\texttt{PoliticalHumor}, 
\texttt{TrumpCriticizesTrump}.
and \texttt{The\_Donald},
In other words, these are the communities that are most likely to be dropped or picked up in the year following the occurrence of opinion change.

The specific subreddits for which behavioral change follows opinion change seem to share a satyrical nature, with `memes' as the main communication medium.
For instance, \texttt{PoliticalHumor} describes itself as ``A subreddit focused on US politics, and the ridiculousness surrounding them'', while \texttt{ABoringDystopia} and \texttt{LateStageCapitalism} poke fun at modern capitalistic society, and \texttt{ENLIGHTENEDCENTRISM} ridicules the ``hypocrisy of the centrist types''.
We test this hypothesis quantitatively in the following section.

\subsection{Behavioral changes and language}
\label{subsec:rq2}

To test our hypothesis about the nature of the subreddits for which the user participation change is significantly different in the \cmvdelta group, we look at the text of the posts within the subreddits.
We leverage from the work presented by \citet{rashkin2017truth} about \emph{satire, hoaxes, and propaganda}.
In order to automatically classify news belonging to these categories, authors develop a measure of `dramatic language'.
This measure detects in fact the use of specific words belonging to lexical categories associated with ``unreliable sources".
We make use of the resources made available by the authors to measure this aspect in our data.\footnote{The lexica can be retrieved at \url{https://hrashkin.github.io/factcheck.html}.}

In order to employ this lexica in our case, we perform the following steps.
We fit \texttt{scikit-learn}'s tf-idf vectorizer on \citeauthor{rashkin2017truth} original dataset, which contains both reliable and unreliable news.
Here, we use the same preprocessing steps and parameterization of the vectorizer that we use on submissions and subreddits text, and described in \Cref{subsec:textual-corpus}.
Then, we use such vectorizer to obtain a representation of the text extracted from each considered subreddit.
Finally, we compute the dramatic language measure for each subreddit by summing up the entries that corresponds to words  included in the lexica associated with unreliable sources.
We normalize this measure by the number of non-zero entries in the vector in order to take into account the different amount of text present in different subreddits.

\begin{figure}
	\centering
	\includegraphics[width=\columnwidth]{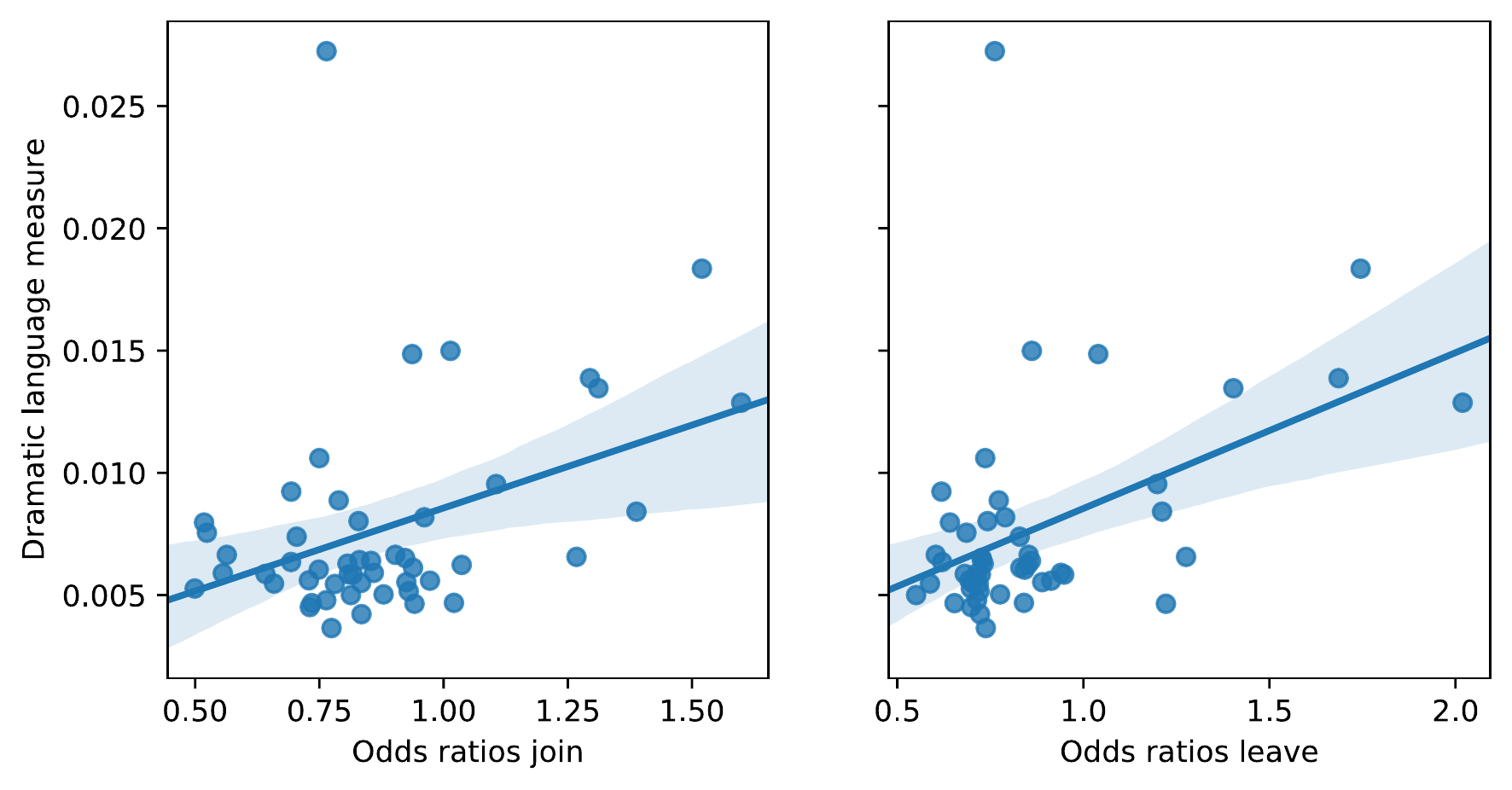}
	\caption{Regression plots which show the correlation between the dramatic language measure and the values of odds ratios $\OR^J_s$ and $\OR^L_s$ for each sociopolitical subreddit $s$.} 
	\label{fig:regplots}
\end{figure}

\Cref{fig:regplots} shows the full scatterplot of the data (one dot for each subreddit) and the linear regression line between the odds ratio and the dramatic language measure.
We find this dramatic language measure to be significantly correlated with the values of odds ratios presented in \Cref{fig:or-delta}, which supports our hypothesis.
As shown in \Cref{tab:dram-lang-corr}, Pearson's correlation coefficient between the odds ratio of each subreddit is around $0.4$ (slightly lower for join, and slightly higher for leave).
In other words, for the \cmvdelta group of users who change opinion, it is more likely that they will leave or join subreddits that make more use of dramatic language---akin to satire, hoaxes, and propaganda.

\begin{table}
	\caption{Pearson's correlation coefficient $r$ and explained variance $R^2$ between odds ratios and dramatic language measure.}
	\label{tab:dram-lang-corr}
	\centering
	\begin{tabular}{l c c}
\toprule
{Event} &       $r$ &  $R^2$ \\
\midrule
Join  &  0.385 $^{**}\phantom{^*}$ &    0.149 $^{**}\phantom{^*}$\\
Leave &  0.458 $^{***}$ &    0.210 $^{***}$\\
\bottomrule
\end{tabular}

\end{table}

\section{Discussion}
\label{sec:discussion}

We have found that opinion change, signaled by leaving an explicit \cmvdelta mark in a discussion on \rcmv, is related to behavioral change, measured as change in user activity on Reddit.
In particular, the text of the submission which marks the opinion change can be used as a predictor for which communities are subject to change, and works as well as the full past user activity profile---a commonly-used predictor in the literature~\cite{massachs2020roots}.
Globally the \cmvdelta group is not any more likely to show behavioral change than the \cmvnodelta group.
Conversely, a large effect on future information consumption behavior is due to posting on \rcmv.
However, there are specific subreddits for which giving a \cmvdelta has an effect (i.e., those in the upper-right cluster in \Cref{fig:or-delta}):
these subreddits are more vulnerable to participation change after opinion change.
They are also associated to a high level of `dramatic language', which is often found in satire, propaganda, and misinformation.
Indeed, most of these communities are satirical in nature, and use `memes' as the main medium of expression.

Starting a discussion on a sociopolitical topic on \rcmv shows an interest in such themes by the user.
In fact, we find that users who post on \rcmv are more likely to change their future participation to Reddit communities than those posting in a generalist community (OR = 1.511).
This effect is larger than the one associated to self-declaring opinion change.
As such, posting on \rcmv can be considered part of a politicization process (in the broadest sense).
And even if the original author of the submission is not politicized a priori, at least part of the comments will be.
Indeed, sociopolitical content is more common on Reddit than one would initially think.
\citet{kane2018communities} show that even in subreddits that have no social or political character, it is possible to recognize traits of politicization in the discussions.
\citet{rajadesingan2021political} report that nearly half of all political talk takes place in subreddits that host political content less than 25\% of the time.
This trend is in line with an increasing politicization of typically non-political spaces that scholars have observed~\citep{dagnes2019us}.
Increasing politicization has also been connected to polarization~\cite{chinn2020politicization}: increasing interest and engagement in politics can increase polarization, while polarization also drives the politicization of non-political spaces.

In this regard, satirical and humorous content (e.g., memes) have been increasingly recognized as essential tools of propaganda in modern politics~\cite{hornback2018afterword}.
Previous research has identified satire, hoaxes, and propaganda as related concepts~\cite{cooke2017posttruth}.
For this reason, they are often clubbed together in some works, including in the one that developed the ``dramatic'' lexicon~\cite{rashkin2017truth} we employ.
In this latter work, authors distinguish these three categories (satire, hoaxes, and propaganda) in their data set, but conclude that the language bears similarities across all of them.
Thus, while the dramatic lexicon is not fine-grained enough to distinguish them, it is in general typical of ``unreliable (esp. hoax and propaganda)'' content.

According to \citet{laaksonen2021clowning}, political humor online has two conflicting effects: on the one hand, it favors engagement with political issues; on the other hand, satire also amplifies polarization and might repel some users.
\citet{penney2020its-so-hard-not} also notes the relationship between political humor online and hyper-polarization.

We found empirical support for these hypothesis in our work.
The usage of satirical and propagandistic language in a community makes it more likely both for it to be left in case of opinion change, but also more likely to be joined (depending on the user).
The increase in odds seems to be strongly correlated between the two cases, thus bolstering the findings from \citet{laaksonen2021clowning} about the ambivalent nature of online political satire.

Different possible explanations can be formulated to interpret this effect.
By building a shared vocabulary and sense of humor, memes and political satire help in delineating in-groups and out-groups~\cite{elgaaied-gambierMeTryingTalk2021,buieWhereYourSense2021}.
The emotional component associated to satire is also a factor that contributes to this conflicting nature.
\citet{utych2018negative} has shown that when people are exposed to political communication expressed with words that have a negative emotional value, they are more likely to reject the content of the message.
This effect can be exemplified by the unpleasant feeling that a meme has on a member of the out-group.
Naturally, this line of reasoning can be applied in general to online propaganda.
Such communities and sources of information are therefore both the first to fall in case of opinion change, as well as the most likely to be joined.

\smallskip

\paragraph{Limitations.}
As with any empirical study, there are some limitations to our work, mostly due the particular dataset used.
First of all, our choice of sociopolitical subreddits could be biased, and many other subreddits might be missing from the selected ones.
However, we adopted a systematic process with precise criteria to select these subreddits in order to minimize bias.
Our choice of subreddits is also consistent (by virtue of the content classifier) with the \rcmvs posts we selected.
Finally, while some subreddits might be missing, adding more subreddits does not invalidate our current analysis.

Another important limitation is that our analysis is observational, and not causal: we do not prove a causal link, but discover a relation between opinion change and a change in information consumption.
However, we do consider the arrow of time in this relation (behavioral changes follow opinion change), and the presence of a topical link (there is a connection between the text of the post and the subreddits that subsequently change).
While limited, this analysis constitutes a necessary first step for any subsequent causal investigation.

Ultimately, our results are necessary limited to the platform we study.
As with any empirical work, it is difficult to prove that our results generalize to different platforms, cultures, or societies.
This limitation stems from the exceptionality of the \rcmvs community, that offers a rich, codified dataset that is not directly available on other mainstream social media.

\smallskip

\paragraph{Contribution and future work.}
To the best of our knowledge, our study is the first step into assessing the relationship between a self-reported opinion change event, and a change in the consumption patterns of online information.
As such, it lays the ground for different lines of research.
Firstly, the characterization of changes in behavior could be further explored, in order to assess which other characteristics of a community or an information source makes it more vulnerable to opinion change.
The measure of dramatic language we use offers a first, coarse-grained characterization of the communities most affected by opinion change events.
Future work includes finding more precise NLP tools to better characterize the behavioral consequences of opinion change, and investigating which users that display more prominent changes in behavior.
This matter has also practical consequences for fighting against misinformation (e.g., on health-related issues) where opinion change, as part of a polarization process, can make users more exposed to hoaxes.
Therefore, it would be helpful to verify our findings in other media and societies.
Furthermore, identifying the effects of opinion change events can also be leveraged to provide empirical bases to recognize them, contributing thus to a more empirical understanding of the processes of opinion dynamics.

\subsection{Ethical Considerations}
As researchers working with user-generated data, our first consideration must be devoted to evaluating whether the data we used in this work was properly collected and treated.
In this study, Reddit users that authored any content we used were in general aware of the public nature and accessibility of their content: the communities under study are in fact in the public domain, visible without any account or password, and have thousands of participants.
We highlight that we did not recover any content willingly deleted by its author.
Furthermore, no personally identifiable information was ever collected: Reddit users make use of pseudonyms, and the messages employed in this study involve only one's view on general, broad topics, making it difficult to uncover any participant's identity.
Finally, all the data was used and presented only as aggregated estimates.

Further considerations must tentatively evaluate the potential broader impact of our work.
Among positive outcomes, identifying the relationship between opinion change, polarization and online information consumption patterns can advance our understanding of how dis- and mis-information sources can be picked up by social media users, and help design public information campaigns.
Of course, however, the same knowledge can also be used to leverage opinion change events in order to spread disinformation.
Subsequently, any ethical judgement on possible applications of our work on information spreading ultimately depends on a value of merit on the information being spread.
However, we believe that understanding how do we change our opinion online and how this affects our information diet helps in different ways.
On the one hand, it recognizes the importance that one's views have on the external world; on the other hand, it furthers our comprehension of how we might fall victims of propaganda.

\bibliographystyle{aaai}
\bibliography{references}

\section{Appendix}
\label{sec:appendix}

\begin{table}[H]
\caption{List of considered sociopolitical subreddits. This list was obtained by manually labeling the \num{2000} most popular subreddits in 2019 as sociopolitical or not, following the definition given by \citet{moy2006predicting}. }
  \label{}
  \begin{tabular}{lll}
    \toprule
    \multicolumn{2}{c}{Sociopolitical subreddits} \\
    \midrule
    ABoringDystopia &
    Anarchism \\
    antiwork &
    AskHistorians \\
    askphilosophy &
    atheism \\
    Bad\_Cop\_No\_Donut &
    Buddhism \\
    Christianity &
    communism \\
    Conservative &
    conspiracy \\
    conspiracytheories &
    CryptoMarkets \\
    Economics &
    economy \\
    ENLIGHTENEDCENTRISM &
    environment \\
    Feminism &
    Firearms \\
    gatekeeping &
    geopolitics \\
    history &
    JordanPeterson \\
    LateStageCapitalism &
    Libertarian \\
    lostgeneration &
    MensRights \\
    Military &
    NeutralPolitics \\
    news &
    OurPresident \\
    philosophy &
    PoliticalDiscussion \\
    PoliticalHumor &
    politics \\
    progun &
    ProtectAndServe \\
    Republican &
    SandersForPresident \\
    skeptic &
    socialism \\
    The\_Donald &
    TheRedPill \\
    TrollXChromosomes &
    TrueReddit \\
    TrumpCriticizesTrump &
    TwoXChromosomes \\
    ukpolitics &
    WitchesVsPatriarchy \\
    worldnews \\
    \bottomrule
  \end{tabular}
\end{table}

\end{document}